\documentclass[onecolumn,aps,prd,preprintnumbers,showpacs,superscriptaddress,nofootinbib,amsmath,amssymb,floats,floatfix,showkeys,notitlepage,longbibliography]{revtex4-1}

\usepackage{comment}
\usepackage{lipsum}
\usepackage{graphicx}
\usepackage{subfigure}
\usepackage{palatino}
\usepackage{sans}
\usepackage{hyperref}
\hypersetup{colorlinks=true,linkcolor=blue,urlcolor=blue,citecolor=blue}
\usepackage[toc,page]{appendix}
\usepackage[normalem]{ulem}
\usepackage{adjustbox}
\usepackage{latexsym}
\usepackage{amsmath}
\usepackage{amssymb}
\usepackage{amsfonts}
\usepackage{dcolumn}
\usepackage{bm}
\usepackage{tikz}
\usepackage{bigints}
\usepackage{array,tabularx,multirow,booktabs}
\usepackage[tracking=true]{microtype}
\usepackage{soul} 
\SetTracking{}{500}
\SetTracking{encoding={*}, shape=sc}{40}
\UseRawInputEncoding 
\allowdisplaybreaks

\begin{document} \sloppy
\title{Apparent and emergent dark matter around a Schwarzschild black hole}

\author{Reggie C. Pantig}
\email{rcpantig@mapua.edu.ph}
\affiliation{Physics Department, Map\'ua University, 658 Muralla St., Intramuros, Manila 1002, Philippines.}

\begin{abstract}
Inspired by the two different dark matter frameworks that were studied recently: one that arises from the non-local effects of entanglement entropy as emergent gravity (characterized by the parameter $\xi(M)$, and zero-point scale length $l$), and one from dark energy viewed as a superconducting medium (characterized by $\eta(M)$, and screening length parameter $\lambda_{\rm G}$), two black hole solutions spherically surrounded with these dark matter models were derived. The effect of these two frameworks on SMBH was analyzed through the resulting deviations in the null regions and the black hole shadow. In addition, constraints to the parameters $\xi$ and $\eta$ (at $3\sigma$ level) were found using the available EHT data for Sgr. A* and M87*. These constraints allow one to deduce the effective mass $M$, which causes uncertainties in the measurement. On the other hand, if the effective mass is known, one can also deduce the constants associated with $\xi$, and $\eta$. The former framework also introduces an Appell function, a hypergeometric function of two variables that separately allows the analysis of macroscopic and (hypothetical) microscopic black holes. This first framework was found to decrease the radii of the null regions respective to the Schwarzschild counterpart. The shadow radius, however, behaves reversibly. The result of the numerical analysis for the latter framework revealed increased values for the photonsphere and shadow radii. Remarkably, the study also showed that for SMBHs, the amplifying effects of $\lambda_{\rm G}$ are stronger than the scale length $l$. Finally, results of constraints, as an example, for the upper bound in $\xi$ for M87* indicated that the effective mass causing the deviation was around $2.4\times 10^{20} M_{\odot}$ given that the observed Milgrom's constant is $5.4\times 10^{-10} \text{ m/s}^2$.
\end{abstract}


\maketitle


\section{Introduction} \label{intr}
Through the vast expanse of the cosmos, two enigmatic entities dominate our understanding of the universe: black holes (BH) and dark matter (DM). Black holes, the remnants of massive stars that have collapsed under their own gravity, continue to captivate the scientific community and the public imagination with their gravitational prowess, capable of bending space and time beyond comprehension. Despite their theoretical prediction by Einstein's general theory of relativity \cite{Schwarzschild:1916uq}, the direct evidence for black holes remained elusive for decades until the pioneering work of astronomers like Andrea Ghez and Reinhard Genzel, who observed stars orbiting an invisible object at the center of our Milky Way, providing compelling evidence for the existence of supermassive black holes \cite{Ghez:2008ms}. On April 10, 2019, the Event Horizon Telescope (EHT) collaboration obtained the first electromagnetic and polarized-based image of the black hole shadow at the heart of Messier 87 \cite{EventHorizonTelescope:2019dse,EventHorizonTelescope:2019ths, EventHorizonTelescope:2022xqj}. Then, the image of the supermassive black hole (SMBH) black hole shadow at the heart of our galaxy, the Milky Way, followed \cite{EventHorizonTelescope:2022wkp,EventHorizonTelescope:2022wok}, confirming the existence of these enigmatic objects. Not only that these results confirm the seminal works by Synge, Luminet, and Falcke \cite{Synge:1966okc, Luminet:1979nyg,Falcke:1999pj}, but they also sparked great interest among the scientific community to study the black hole shadow in various black hole models coming from different phenomenologies and alternative theories of gravity (See Refs. \cite{Bambi:2019tjh,Zahid:2022eeq,Rayimbaev:2022znx,Mustafa:2022xod,Afrin:2022ztr,Zubair:2023cep,Zahid:2023csk,Ghorani:2023hkm,Sekhmani:2024fjn,Raza:2023vkn} to refer a few).

In parallel, the cosmic census has revealed an astonishing truth: the visible matter that we are made of - stars, planets, galaxies - comprises only a fraction of the total matter content of the universe. The majority, approximately 85\% \cite{deVries:1999tiy}, is concealed within the shroud of darkness known as dark matter. Yet, despite its pervasive presence, dark matter remains elusive, evading detection through conventional means due to its lack of interaction with electromagnetic radiation. Various theoretical models have been proposed in the quest to unravel the enigma of dark matter, each offering unique insights into its composition and properties (For an excellent and comprehensive reviews, see Refs. \cite{Garrett:2010hd,Urena-Lopez:2019kud,Arun:2017uaw,Arbey:2021gdg}). Associated with the $\Lambda$CDM model is the detection of the Weakly Interacting Massive Particles (WIMP) or axions \cite{deSwart:2017heh}. Scientists have devised ingenious detection methods to probe the mysteries of dark matter, ranging from direct searches using underground detectors \cite{Baum:2018tfw} to indirect approaches through astronomical observations and particle accelerators. However, despite decades of dedicated efforts that sometimes seemingly show positive results \cite{DAMA:2008jlt,Bernabei:2013xsa,Bernabei:2018jrt}, the elusive nature of dark matter persists, debunking claims for detecting it \cite{CRESST:2015txj,PICO:2017tgi,LZ:2022ufs}, and continues to challenge our fundamental understanding of the cosmos. Not only these, but the model is plagued with certain anomalies at the astrophysical level \cite{Moore:1994yx,Klypin:1999uc,Boylan-Kolchin:2011qkt}.

Recent advancements in astrophysical instrumentation, theoretical frameworks, and phenomenologies have unveiled a tantalizing possibility: an SMBH acting as a laboratory to detect unique imprints of dark matter that surrounds it. In the literature, several researchers have attempted to go in this direction, particularly studying DM effects on the black hole shadow \cite{Hou:2018bar,Hou:2018avu,Konoplya:2019sns,Jusufi:2020cpn,Nampalliwar:2021tyz,Pantig:2022whj,Jusufi:2020zln,Jusufi:2022jxu,Pantig:2022sjb,Anjum:2023axh,Ovgun:2023wmc,Capozziello:2023rfv,Capozziello:2023tbo,Yang:2023tip,Gomez:2024ack,Wu:2024hxr}, gravitational lensing effects \cite{Haroon:2018ryd,Pantig:2022toh,Atamurotov:2021hck,Pantig:2021zqe,Qiao:2022nic,Liu:2023xtb,Macedo:2024qky}, null structures such as event horizon and ergosphere \cite{Xu:2018wow,Xu:2020jpv,Xu:2021dkv}, quasinormal modes and greybody factors \cite{Konoplya:2021ube}, thermodynamic properties \cite{Konoplya:2021ube,Errehymy:2023xpc,Rakhimova:2023rie, Sekhmani:2024xyd}, spherical accretions \cite{Saurabh:2020zqg}, super-radiant instabilities \cite{Liu:2022ygf}, wormhole solutions and their analysis \cite{Mustafa:2022fxn,Mustafa:2023qxf}, and others \cite{DellaMonica:2023dcw,Qiao:2024ehj,Konoplya:2022hbl, Zhou:2022eft,Rayimbaev:2021kjs}. Indeed, when more sophisticated astrophysical instrumentation becomes available in the future, black holes may potentially unveil the true nature of dark matter.

The aim of this paper is to derive new black hole solutions surrounded by two recent dark matter models studied in Refs. \cite{Jusufi:2023ayv,Inan:2024noy}. In the first paper, the authors took into account the zero-point length correction $l$ \cite{Nicolini:2019irw} to the gravitational potential $\Phi$, which allows the modification of the apparent DM hypothesis and Newton’s law of gravity. As a result, it was seen that there is some correspondence between Verlinde's gravity \cite{Verlinde:2010hp,Verlinde:2016toy} and the non-local gravity theories. In the second paper, the authors explored the phenomenological aspects of dark energy modeled as an effective superconducting medium, inspired by the quantum gravity point of view, where gravitons are also given some mass. As a result, Newtonian gravity is affected by some screening length parameter $\lambda_{\rm G}$ due to dark energy, leading to the emergence of dark matter. Combining these two dark matter models to the black hole spacetime geometry, we can gain insights into how these apparent and emergent dark matter leaves traces of their effect on the black holes, especially SMBH at the center of galaxies.

The work follows this program: In Sections \ref{sec2} and \ref{sec3}, we derive the two black hole solutions surrounded by dark matter. In Section \ref{sec4}, we study, analyze, and discuss the dark matter effects on photonsphere and shadow. We also attempt to find constraints to dark matter parameters using the EHT data. In Section \ref{conc}, we state conclusive remarks and possible research prospects. In this paper, we used geometrized units by setting $G = c = 1$ and the metric signature $(-,+,+,+)$.

\section{Black hole metric with apparent dark matter arising from the non-locality of emergent gravity} \label{sec2}
Observational evidence strongly suggests that the speed at which stars and other objects rotate within spiral galaxies is directly related to their distance from the galactic center. This relationship holds until you reach the outer regions of the galaxy, where velocities tend to level off, remaining relatively constant even at significant distances from the center. While Newtonian gravity explains this behavior within the galaxy, it falls short when trying to account for rotational patterns outside of it. According to Newton's law of gravitation, objects farther from the galaxy's core should move more slowly, contrary to observations showing a consistent and steady speed at these distances. It's widely accepted that the visible matter within galaxies lacks the gravitational pull necessary to explain these dynamics. The paper in Ref. \cite{Jusufi:2023ayv} addresses this discrepancy using one of the hypotheses of emergent gravity theory - apparent dark matter, where it is found that the astrophysical dark matter density profile is exactly given by
\begin{equation}
    \rho(r) = \frac{\xi  \left(6 l^{4}+5 l^{2} r^{2}+2 r^{4}\right)}{8 r^{\frac{3}{2}} \left(l^{2}+r^{2}\right)^{\frac{7}{4}} \sqrt{2 l^{2}+r^{2}}\, \pi},
\end{equation}
where $\xi = \sqrt{(a_0/6)GM}$ - a quantity that plays an important role far from the galactic center, and $l$ is a length scale parameter comparable to the Planck length $l_{\rm Pl}$. Also in $\xi$, $M$ is the baryonic mass, and $a_0 = 5.4\times10^{-10} \text{ m/s}^2$ which is a constant found by Milgrom \cite{Milgrom:1983ca} in 1983. The exact dark matter mass profile can then be found as
\begin{equation} \label{mprofile_exact}
    M_{\rm D}= 4 \pi \int_{0}^{r} \rho\left(r^{\prime}\right) r^{\prime 2} d r^{\prime} = \frac{\xi  \,r^{\frac{3}{2}} \sqrt{2 l^{2}+r^{2}}}{\left(l^{2}+r^{2}\right)^{\frac{3}{4}}}.
\end{equation}
Since the length scale $l$ is very small, we can approximate the dark matter mass as
\begin{equation} \label{mprofile_approx}
    M_{\rm D} \simeq \xi r \left( 1 + \frac{l^2}{4r^2} \right) - \mathcal{O}(l^4).
\end{equation}
Since the aim is to envelop a Schwarzschild black hole with a spherical dark matter halo, it is very useful to express any variable that has a dimension of length as a fraction of the black hole mass $m$. In doing so, these variables will become dimensionless. Such a procedure would also mean that we set $m=1$. For instance, we can define $\tilde{r} = r/m$, but we could also say $\tilde{r} \rightarrow r$, where the consensus exists that $r$ is dimensionless. From here on, we keep this in mind.

For any test particle influenced by this mass profile, the tangential velocity can be calculated easily as
\begin{equation}
    v_{\rm tg}(r) = \sqrt{\frac{M_{\rm D}}{r}} \simeq \frac{\xi(4r^2-l^2)}{16\pi r^4}.
\end{equation}
Consider the line element of a spherical dark matter halo
\begin{equation} \label{halo_metric}
    ds^{2}_{\rm DM} = -a(r) dt^{2} + b(r)^{-1} dr^{2} + r^2 d\theta ^{2} +r^2\sin^2\theta d\phi^{2}.
\end{equation}
We can relate $a(r)$ and $v_{\rm tg}(r)$ through \cite{Xu:2018wow}
\begin{equation} \label{vtg}
    v_{\rm tg}(r)=r \frac{d \ln (\sqrt{a(r)})}{d r}.
\end{equation}
After integrating using the mass profile in Eq. \eqref{mprofile_exact},
\begin{equation} \label{a(r)}
    a(r) = 4\xi \sqrt{\frac{2r}{l}} \rm{AppellF}_1\left(\frac{1}{4};\frac{3}{4},-\frac{1}{2};\frac{5}{4};-\frac{r^2}{l^2},-\frac{r^2}{2l^2} \right),
\end{equation}
where we see the appearance of an Appell hypergeometric function of two variables \cite{appell1926fonctions, srivastava1985multiple}. Since we initially defined $l$ to be very small, using Eq. \eqref{mprofile_approx} to Eq. \eqref{vtg} yields
\begin{equation}
    a(r) = 2\xi \ln(r) -\frac{\xi l^2}{4r^2},
\end{equation}
which is only applicable to astrophysical applications to supermassive black holes.

From here on, embedding the metric of the halo to the black hole spacetime metric can be done by following certain procedures \cite{Xu:2018wow}. To review briefly, the idea is to write the metric functions as
\begin{equation}
    A(r)=a(r) + F(r), \quad \quad B(r) = b(r)+G(r),
\end{equation}
to the line element
\begin{equation} \label{metric1}
    ds^{2} = -A(r) dt^{2} + B(r)^{-1} dr^{2} + r^2 d\theta ^{2} +r^2\sin^2\theta d\phi^{2}.
\end{equation}
Solving the Einstein field equation
\begin{equation}
    R^{\mu}_{\nu}=\frac{1}{2}\delta^{\mu}_{\nu}R=\kappa^2((T^{\mu}_{\nu})_{\rm DM}+(T^{\mu}_{\nu})_{\rm Schw}),
\end{equation}
gives the relation
\begin{align}
    (b(r)+G(r))\left(\frac{1}{r^{2}}+\frac{1}{r}\frac{b'(r)+G'(r)}{b(r)+G(r)}\right)&=b(r)\left(\frac{1}{r^{2}}+\frac{1}{r}\frac{b'(r)}{b(r)}\right), \nonumber \\
    (b(r)+G(r))\left(\frac{1}{r^{2}}+\frac{1}{r}\frac{a'(r)+F'(r)}{a(r)+F(r)}\right)&=b(r)\left(\frac{1}{r^{2}}+\frac{1}{r}\frac{a'(r)}{a(r)}\right).
\end{align}
After solving for $F(r)$ and $G(r)$, we get
\begin{equation} \label{e19}
    A(r) = \exp\left[\int \frac{b(r)}{b(r)-\frac{2}{r}}\left(\frac{1}{r}+\frac{a'(r)}{a(r)}\right)dr-\frac{1}{r} dr\right], \quad \quad B(r) =b(r)-\frac{2}{r},
\end{equation}
It is easy to see that without the dark matter halo, $a(r) = g(r) = 1$, and the integral in $A(r)$ results in a constant equal to $1 - 2/r$. Having a conservative approach, where we surround a Schwarzschild black hole with the halo resulting from the non-locality of the emergent gravity theory, the assumption that $a(r)=b(r)$ and $F(r)=G(r)=-2/r$, it implies that $A(r)=B(r)$, giving a metric function expressed as
\begin{equation} \label{app_metric}
    A(r) \simeq r^{2 \xi}-\frac{r^{2 (\xi -1)} \xi  \,l^{2}}{4} - \frac{2}{r} + \mathcal{O}(l^4).
\end{equation}

\section{Black hole metric with emergent dark matter from superconducting dark energy} \label{sec3}
Another model we can consider in this paper is the emergent property of dark matter arising from a graviton mass due to dark energy mimicking a superconductor \cite{Inan:2024noy}, which has been considered recently. It has been shown that Newtonian gravity, due to the presence of dark energy viewed similar to the properties of a superconductor, yields a screening length parameter
\begin{equation}
    \lambda_{\rm G} = \frac{1}{\sqrt{2\Lambda}},
\end{equation}
where $\Lambda$ is the cosmological constant. The phenomenological treatment in \cite{Inan:2024noy} has shown the emergence of the Yukawa-like gravitational potential
\begin{equation}
    \Phi_{\rm DM}(r) = \frac{\mathcal{C}}{r}e^{\frac{-r}{\lambda_{\rm G}}},
\end{equation}
where $\mathcal{C} = -\alpha M \mu$. Here, $\alpha$ is a dimensionless parameter which encodes correlations between the matter field and gravitons. Moreover, $M$ is the baryonic mass, while $\mu$ is the test particle's mass. In galactic dynamics, the dark matter mass was shown to be expressed as \cite{Inan:2024noy}
\begin{equation}
    M_{\rm D}(r)= \eta \left( \frac{r + \lambda_{\rm G}}{\lambda_{\rm G}}e^{-\frac{r}{\lambda_{\rm G}}} \right),
\end{equation}
where we have written $\eta = \alpha M$. The rotational velocity due this dark matter distribution is then
\begin{equation}
    v_{\rm tg} = \sqrt{\frac{\eta}{r}\left( \frac{r + \lambda_{\rm G}}{\lambda_{\rm G}}e^{-\frac{r}{\lambda_{\rm G}}} \right)}.
\end{equation}
Using Eq. \eqref{vtg} and upon integrating,
\begin{equation}
    a(r) = -\frac{2\eta}{r}e^{-\frac{r}{\lambda_{\rm G}}}.
\end{equation}
Following the procedures on Sect. \ref{sec2}, the metric function of a black hole around an emergent dark matter is given by
\begin{equation} \label{emer_metric}
    A(r) = \exp{\left[ -\frac{2\eta}{r}e^{-\frac{r}{\lambda_{\rm G}}} \right]} - \frac{2}{r}.
\end{equation}

\begin{figure*}
    \centering
    \includegraphics[width=0.48\textwidth]{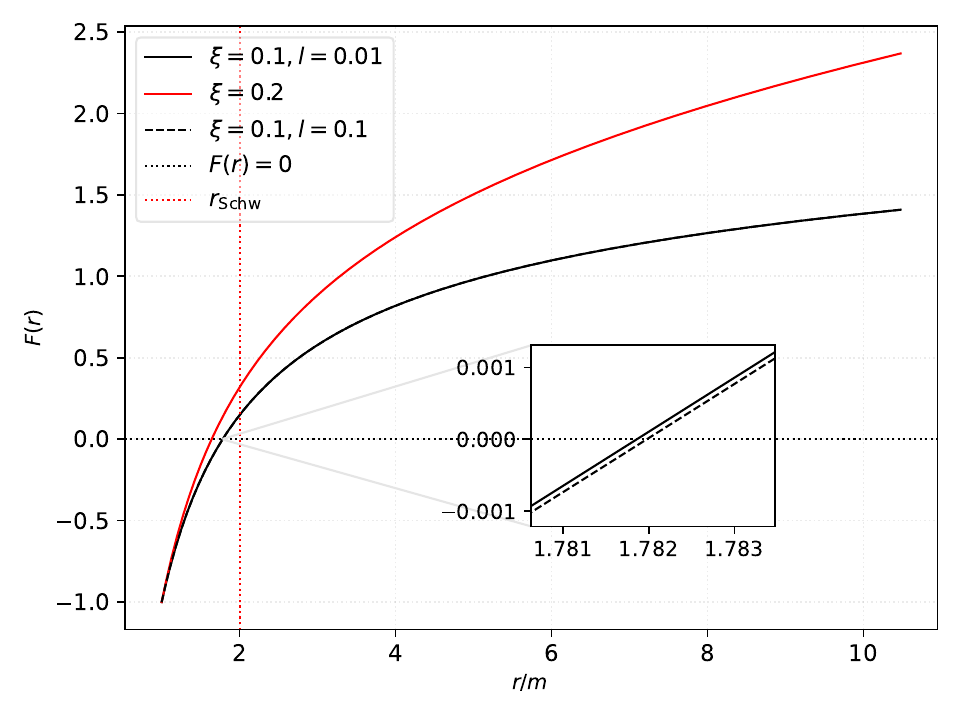}
    \includegraphics[width=0.48\textwidth]{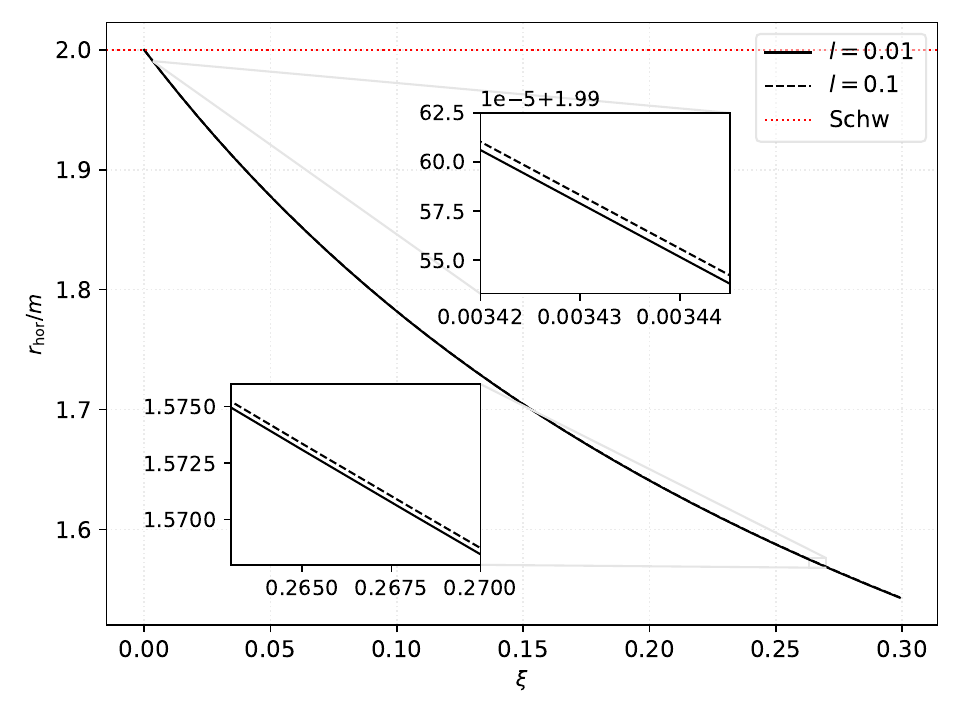}
    \includegraphics[width=0.48\textwidth]{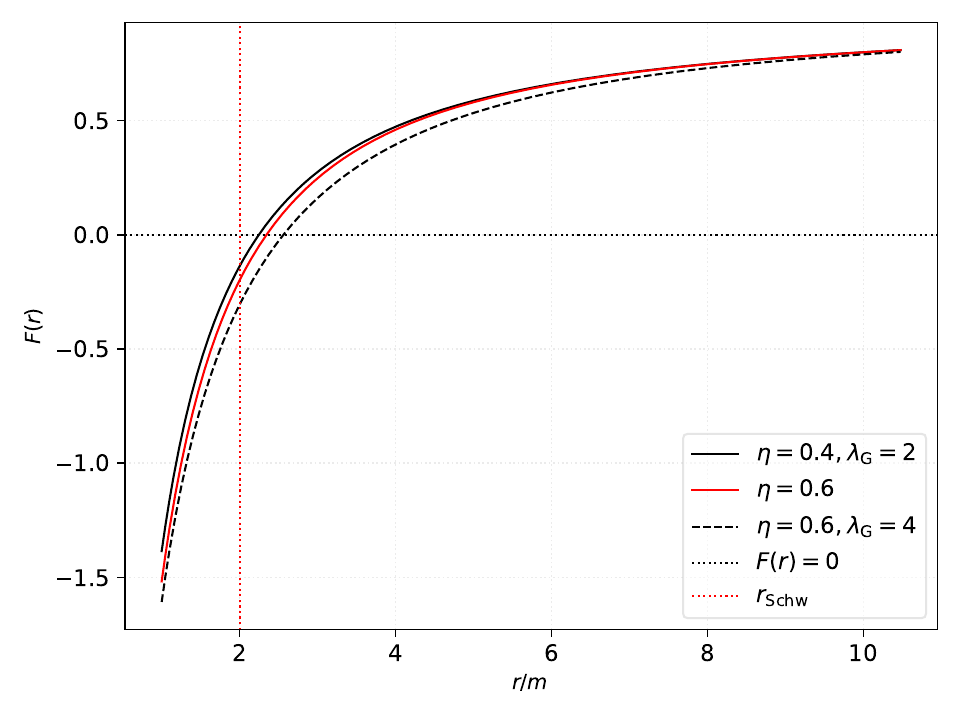}
    \includegraphics[width=0.48\textwidth]{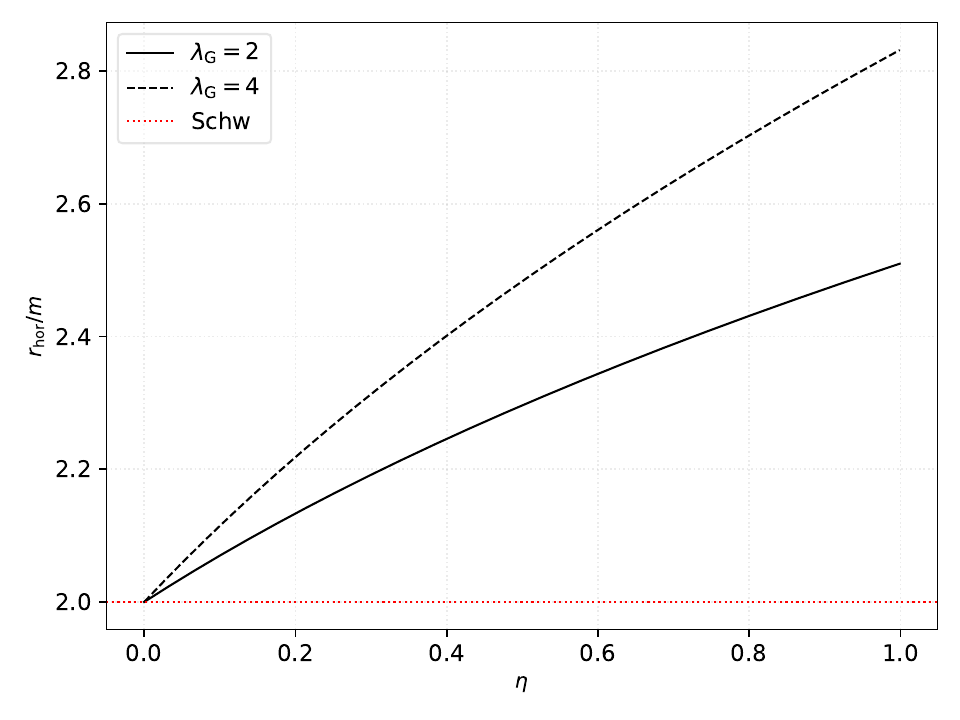}
    \caption{Deviation of the event horizon due to the non-locality of emergent gravity, plotted in two different ways. We include the Schwarzschild case for comparison (vertical and horizontal red dotted lines in the upper left and right panels, respectively). The lower panels consider the horizon effects of emergent dark matter due to dark energy viewed as a superconductor.}
    \label{hor}
\end{figure*}

 How the horizon radius behaves under the influence of the parameters, $\xi$ and $\eta$ are plotted in Fig. \ref{hor}. It only shows how the different dark matter models considered in this study alter the horizon of a Schwarzschild black hole. If we base on the non-locality feature of emergent gravity, the horizon shrinks as $\xi$ increases. Interestingly, the included effect of the length scale $l$ is shown on the inset plot, where it is seen to increase the horizon radius as its value increases. We take note, however, that for astrophysical black holes such as SMBHs, $l$ is vanishingly small, and its effect can be neglected. In the upper right panel, we observe that the deviation between two values of $l$ is large for the case where the dark matter density is high (large $xi$).
 
 Meanwhile, if dark matter emerges based on the view of dark energy mimicking a superconductor, the horizon radius increases with $\eta$. The screening length parameter further enhances this increase (see the lower left panel). Note that $\eta$ is related to the product of the constant $a_0$ and $M$. On the right panel, the change in the horizon radius given two values of the screening length parameter is drastically increasing given more dark matter mass.

For the next section, we can recast the line element in Eq. \eqref{metric1} as
\begin{equation} \label{metric2}
    ds^{2} = -A(r) dt^{2} + B(r) dr^{2} + C(r) d\theta ^{2} +D(r) d\phi^{2},
\end{equation}
where $B(r)=A(r)^{-1}$, $C(r)=r^2$, and $D(r)=r^2\sin^2\theta$. The conditions  $(\partial_t g_{\mu\nu} = 0), (\partial_\theta g_{\mu\nu} = 0, \partial_\phi g_{\mu\nu} = 0)$ implies that the spacetime is static and spherically symmetric. Throughout the paper, we specialize at $\theta = \pi/2$, allowing us to simplify the analysis through a $1+2$ dimensions only: $ds^{2} = -A(r) dt^{2} + B(r) dr^{2} + r^2 d\phi^{2}$.

\section{Null geodesic, black hole shadow and EHT constraints} \label{sec4}
\begin{figure*}
    \centering
    \includegraphics[width=0.48\textwidth]{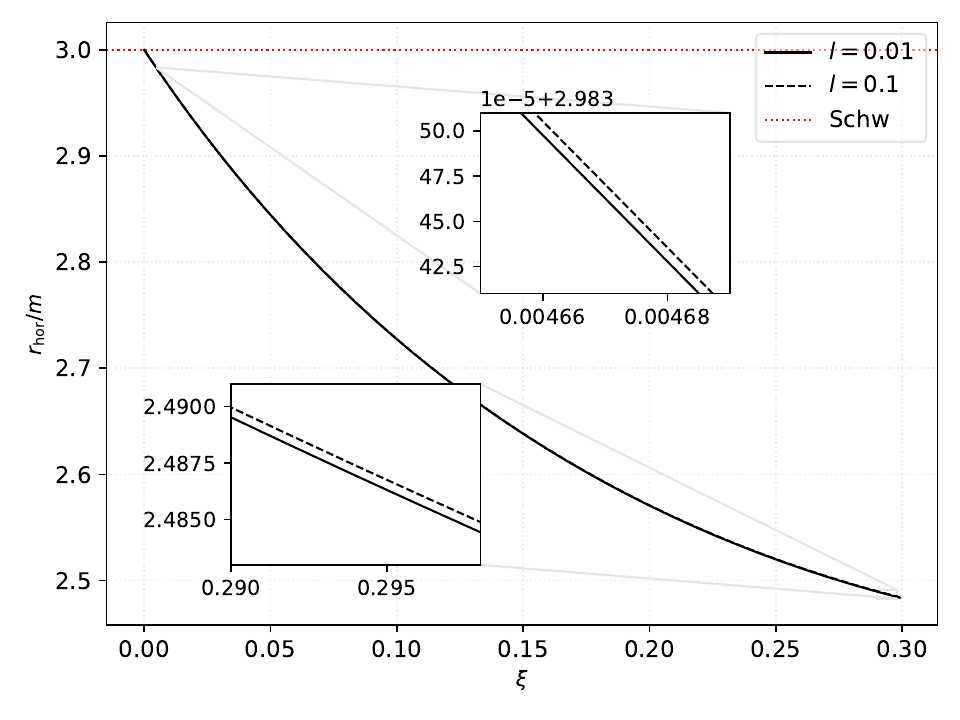}
    \includegraphics[width=0.48\textwidth]{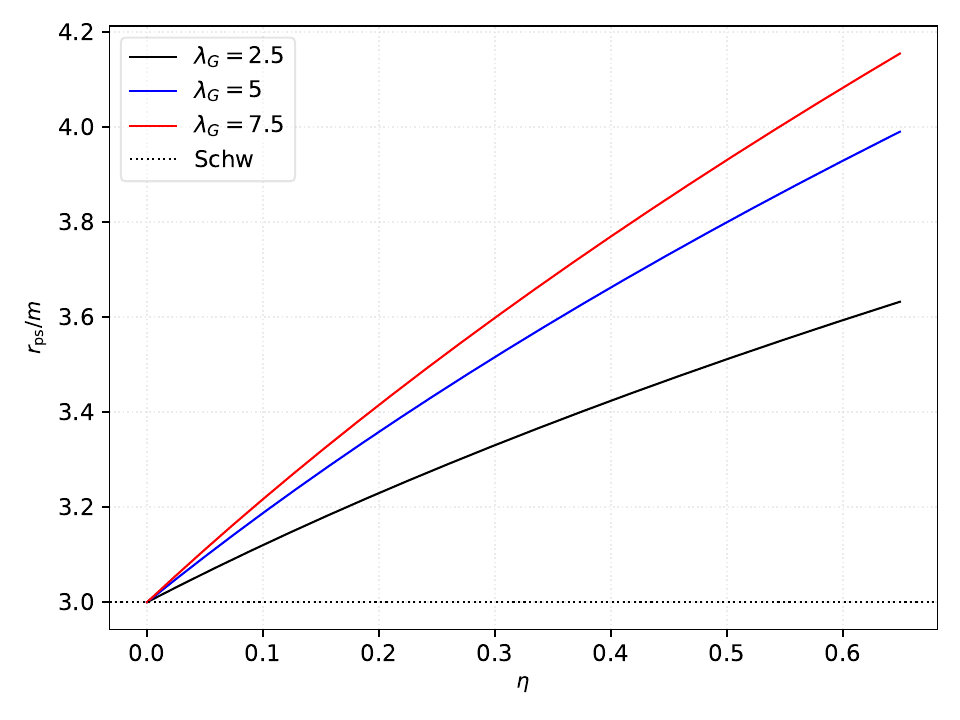}
    \caption{Deviation of the photonsphere radius due to the non-local effects of emergent gravity (left panel), and due to dark energy viewed as a superconductor (right panel). The horizontal dotted line is the Schwarzschild photonsphere radius, which serves as a comparison.}
    \label{rps}
\end{figure*}

We first analyze how the photonsphere will be affected by the corresponding dark matter profiles that we considered. It is well-known that this analysis can be achieved by considering the Lagrangian 
\begin{equation}
    \mathcal{L} = \frac{1}{2} g_{\mu\nu} \dot{x}^\mu \dot{x}^\nu = \frac{1}{2}\left( -A(r) \dot{t}^2 + B(r) \dot{r}^2  + C(r) \dot{\phi}^2 \right),
\end{equation}
which leads to the two constant of motion:
\begin{equation}
    E = A(r)\frac{dt}{d\lambda}, \qquad L = C(r)\frac{d\phi}{d\lambda}.
\end{equation}
Their ratio defines the impact parameter of light rays
\begin{equation}
    b =\frac{L}{E} = \frac{C(r)}{A(r)}\frac{d\phi}{dt},
\end{equation}
which involves the rate of change of $\phi$. In analyzing null geodesics, $ds^2 =0,$ implying that $g_{\mu \nu}\dot{x}^\mu \dot{x}^\nu = 0$, which allows one to derive the orbit equation. That is, how the radial coordinate $r$ change with $\phi$:
\begin{equation}
    \left(\frac{dr}{d\phi}\right)^2 =\frac{C(r)}{B(r)}\left(\frac{h(r)^2}{b^2}-1\right),
\end{equation}
where the function $h(r)^2$ is defined as
\begin{equation}
    h(r)^2 = \frac{C(r)}{A(r)}.
\end{equation}
From the orbit equation, the two conditions $dr/d\phi = 0$ and $d^2r/d\phi^2 =0$ implies that the photonsphere radius can be found through $h'(r) = 0$. In other words, we solve $r$ from the expression
\begin{equation}
    A(r)C'(r) - A'(r)C(r) = 0.
\end{equation}
Depending on how complicated $A(r)$ and $C(r)$ is, the solution to $r \rightarrow r_{\rm ps}$ can be found either analytically or numerically. When we use Eq. \eqref{app_metric}, the location of the photonsphere can be found by solving $r$ in
\begin{equation} \label{rph1}
    r^{2\xi - 1}\left[\frac{l^{2} \xi  \left(\xi -2\right)}{2}-2(\xi -1) \right] - 6 = 0.
\end{equation}
For Eq. \eqref{emer_metric}, we find
\begin{equation} \label{rph2}
    -\eta  \left(\lambda_{\rm G} +r \right) {\mathrm e}^{\frac{-2 \eta  \,{\mathrm e}^{-r /\lambda_{\rm G}} \lambda_{\rm G} -r^{2}}{\lambda_{\rm G}  r}} + \lambda_{\rm G} \left(r{\mathrm e}^{-\frac{2 \eta  \,{\mathrm e}^{-r/\lambda_{\rm G}}}{r}} -3 \right) = 0.
\end{equation}
It can be seen how these expressions simply reduce to $r_{\rm ps}/m = 3$ where dark matter is not present. Theoretical results are shown in Fig. \ref{rps}.
In the left panel, $\xi$ has shown more effect to $r_{\rm ps}$ than the scale length $l$. If only $l$ is large, it can be seen how it increases the photonsphere by a minuscule amount. Since $l \sim 0$, the curve in the plot can be safely attributed to $\xi$ alone. What we could also derive from the analysis is that as $\xi$ gets larger, the more it amplifies the effect of the $l$. In the right panel, both the parameters $\lambda_{\rm G}$ and $\eta$ increases the photonshere radius. The only difference between emergent dark matter from dark energy to the one coming from the non-locality of the emergent gravity is that the $\eta$ is shown to have more amplifying effects on the length parameter $\lambda_{\rm G}$.

\begin{figure*}
    \centering
    \includegraphics[width=0.48\textwidth]{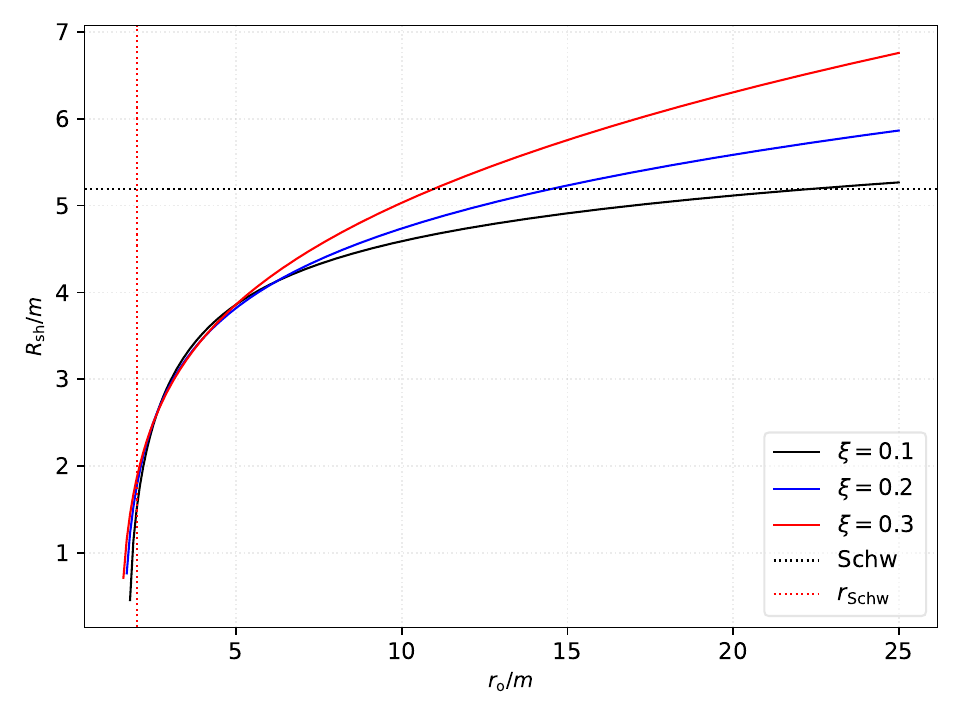}
    \includegraphics[width=0.48\textwidth]{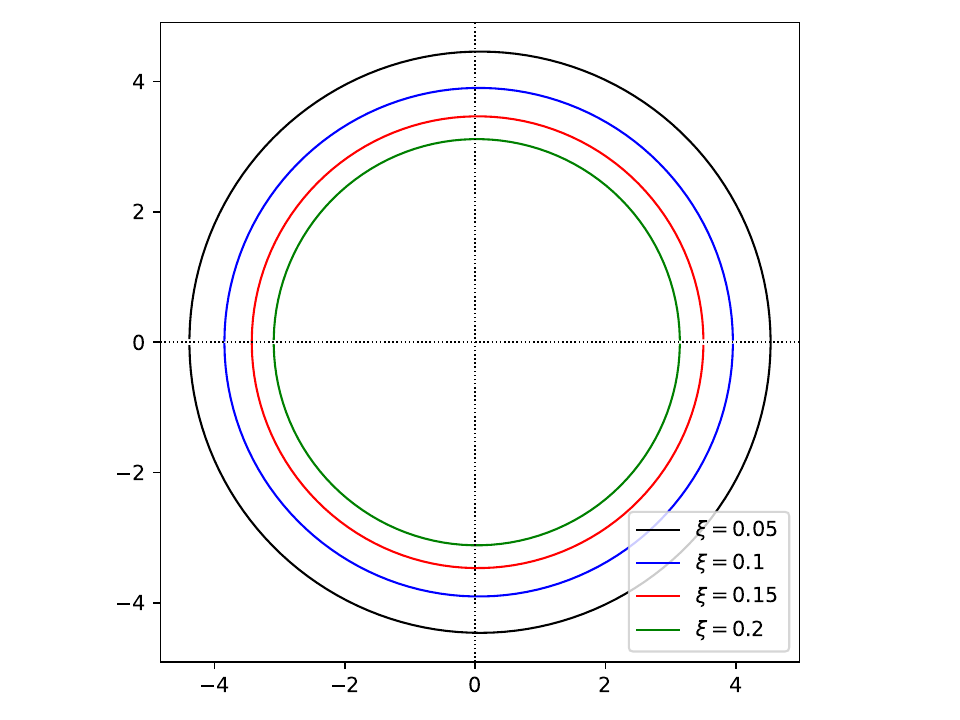}
    \includegraphics[width=0.48\textwidth]{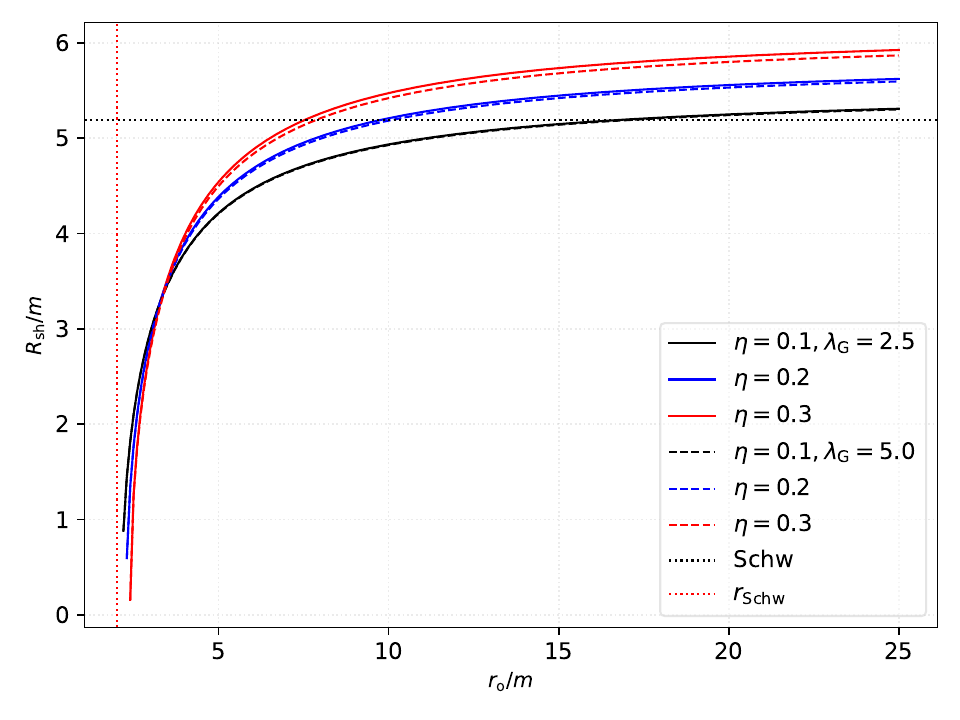}
    \includegraphics[width=0.48\textwidth]{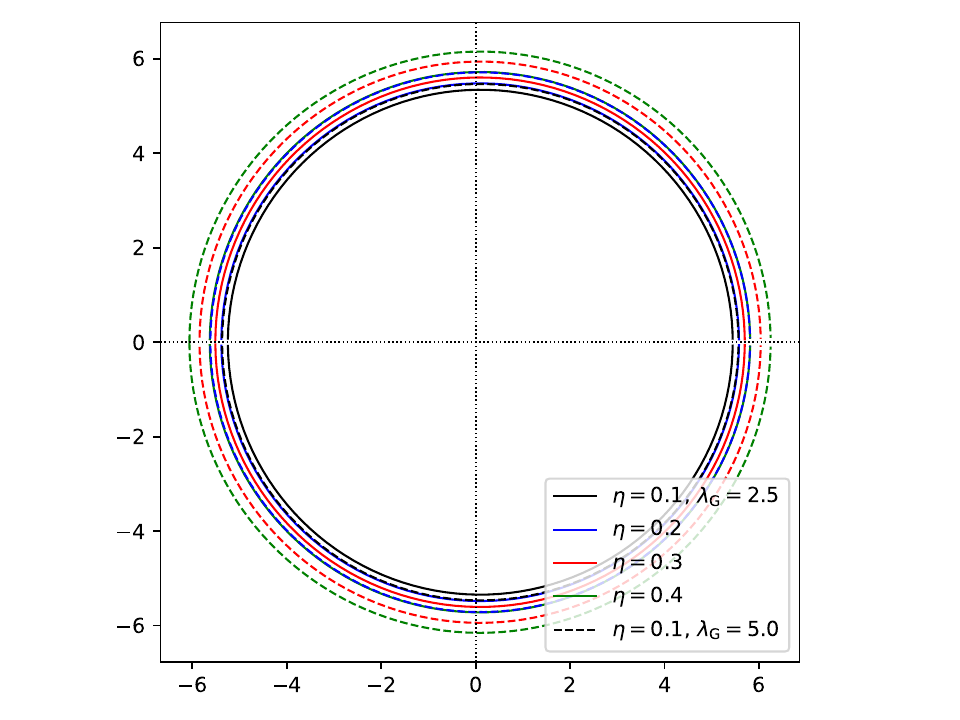}
    \caption{The figures on the left panel show the theoretical behavior of the shadow radius that is observer-dependent. The vertical red and horizontal black dotted lines correspond to the Schwarzschild horizon and shadow. The right panel shows the shadow cast as perceived by some remote observer.}
    \label{rsh_theo}
\end{figure*}

The circular orbit of photons is inherently unstable. That is, any small perturbation will dictate whether it will spiral into the black hole or escape. Hence, backward tracing these escaping photons from the photonsphere will travel along the intervening space under the influence of dark matter. Originating from the photonsphere, these photons will travel to the observer at the critical impact parameter $b_{\rm crit}$. Thus, we can see how the photonsphere and the impact parameter are related to the shadow's perceived angular radius and radius. Following the procedures developed in Ref. \cite{Perlick:2021aok}, an observer-dependent angular shadow reads
\begin{equation} \label{angl_sha}
    \sin^{2}(\Theta_{\rm sh}) = \frac{b_{\rm crit}^{2}}{h(r_{\rm o})^{2}},
\end{equation}
where $r_{\rm o}$ is the radial location of an observer, and the critical angle as
\begin{equation}
    b_{\rm crit} = \sqrt{\frac{C(r_{\rm ps})}{A(r_{\rm ps})}}.
\end{equation}
With the values for $r$, which represents the photonsphere, that we obtained from the numerical analysis of Eqs \eqref{rph1}-\eqref{rph2}, the shadow radius can be calculated as
\begin{equation} \label{eqshad}
    R_{\rm sh} = \sin(\Theta_{\rm sh}) r_{\rm o} = b_{\rm crit}\sqrt{A(r_{\rm o})}.
\end{equation}
We examine plots derived from theoretical considerations, shown in the left panel of Fig. \ref{rsh_theo}. In these plots, the observer-dependence of the shadow radius is also evident. Generally, the shadow radius is small when the observer is near the black hole.  The upper left panel shows $R_{\rm sh}$ curve under the influence of the non-locality of emergent gravity. An observer around the range $r_{\rm o}$ may perceive a larger shadow for $\xi = 0.1$ than $\xi = 0.3$. As the observer goes very far from the black hole, the effect reverses. Higher values of $\xi$ make the observer perceive the Schwarzschild shadow at a closer distance from the black hole. Interestingly, while the photonsphere radius decreases while $\xi$ increases, the reverse effect happens to the shadow radius. The parameter $\xi$ has been shown to amplify the difference between the shadow radius from $\Delta \xi = 0.1$. The shadow cast for this case is shown in the upper right panel, plotted once the numerical value for $R_{\rm sh}$ is found as we approximate $r_{\rm o} \rightarrow \infty$.

The emergent dark matter arising from dark energy modeled as a superconductor has shown a different behavior for the shadow radius. See the lower left panel. First, both $r_{\rm ps}$ and $R_{\rm sh}$ increases with $\eta$. Comparing this to the apparent dark matter from the non-local emergent gravity model, $\eta$ does not show any amplification between two values of $R_{\rm sh}$ even when the observer distance from the black hole is large, which is a different behavior than the one we observe for the photonsphere. Interestingly enough, for a given value of $\eta$, increasing the value of the screening length parameter increases $r_{\rm ps}$ but is shown to decrease $R_{\rm sh}$. The shadow cast plotted in the lower right panel agrees with the observation from the remote observer.
 
\begin{figure*}
    \centering
    \includegraphics[width=0.48\textwidth]{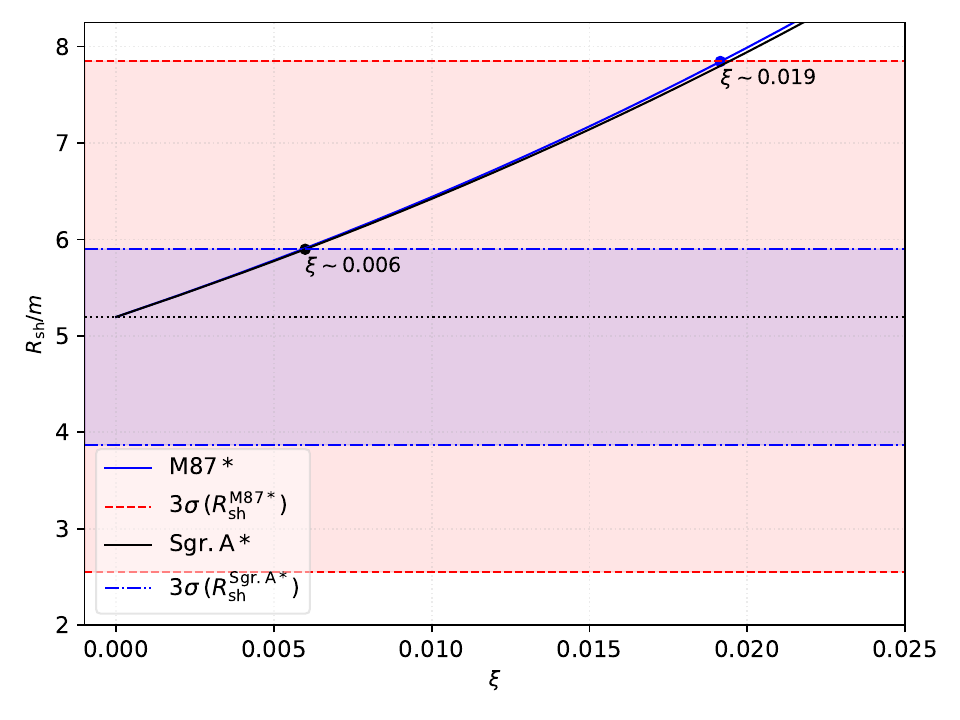}
    \includegraphics[width=0.48\textwidth]{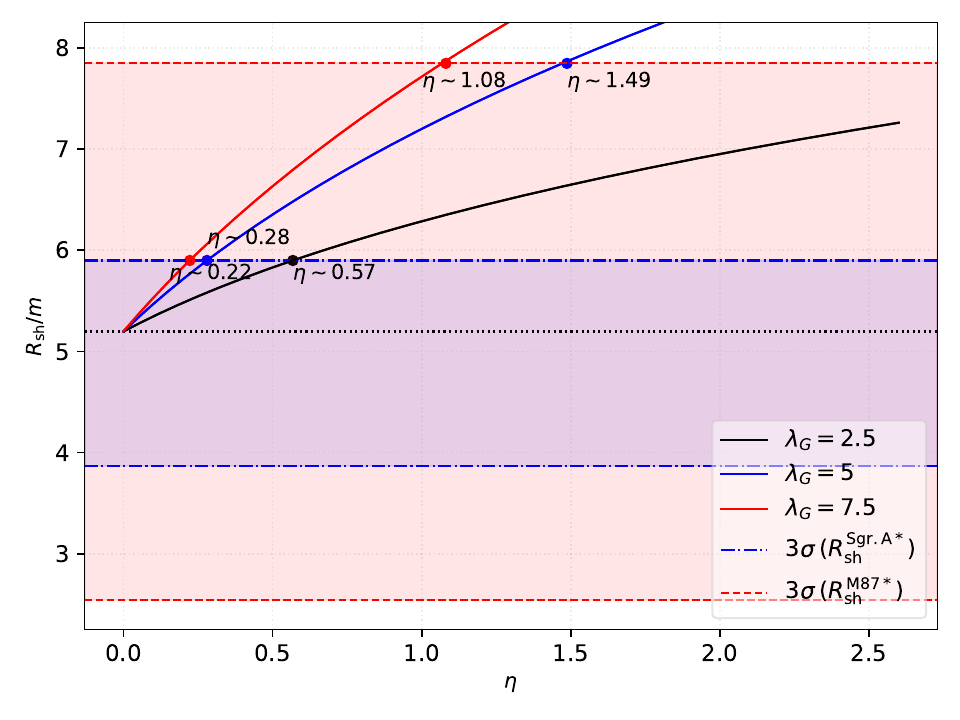}
    \caption{Constraints on $\xi$ and $\eta$ using actual data from EHT for Sgr. A* (blue shade) and M87* (red shade). The annotated values indicated are the upper bounds. The lower bound is always zero, representing the Schwarzschild case not surrounded by dark matter.}
    \label{rsh}
\end{figure*}

Further numerical analysis is plotted in the right panels of Fig. \ref{rsh} using real data from the EHT as shown in Table \ref{tab1}.
\begin{table}
    \centering
    \begin{tabular}{ p{2cm}  p{3.5cm}  p{4.5cm}  p{2cm}}
    \hline
    \hline
    Black hole & Mass $m$ ($M_\odot$) & Angular diameter: $2\alpha_\text{sh}$ ($\mu$as) & Distance (kpc) \\
    \hline
    Sgr. A*   & $4.3 \pm 0.013$x$10^6$ (VLTI)    & $48.7 \pm 7$ (EHT) &   $8.277 \pm 0.033$ \\
    M87* &   $6.5 \pm 0.90$x$10^9$  & $42 \pm 3$   & $16800$ \\
    \hline
    \end{tabular}
    \caption{EHT data for Sgr. A* and M87*.} \label{tab1}
\end{table}
The aim is to apply a simple parameter estimation through a fitting procedure to validate the models used in this study. We want to see the bounds for $\xi$ and $\eta$, if they exist, at $3\sigma \,(99.7\%)$ confidence level. Looking at Refs. \cite{EventHorizonTelescope:2019dse,EventHorizonTelescope:2022wkp,EventHorizonTelescope:2021dqv,Vagnozzi:2022moj}, the $3\sigma$ level of significance can easily be calculated as $3.871M \leq R_\text{sh} \leq 5.898M$, and $ 2.546M \leq R_\text{sh} \leq 7.846M$ for Sgr. A* and M87*, respectively. In addition to how the shadow radius behaves with $\xi$ and $\eta$, the plot also identifies (see the annotations) the upper bounds of $\xi$ and $\eta$ for Sgr. A* and M87* data. We should note that the lower bound for both is zero, which is the Schwarzschild case.

A final caveat: while actual black holes have spin parameters described by the parameter $a$, constraining $\xi$ and $\eta$ using the non-rotating solution might give unrealistic results. However, it was explained in Ref. \cite{Vagnozzi:2022moj} why considering the non-rotating case is sufficient in finding constraints for parameters affecting the black hole geometry. Furthermore, we also neglect radiation escaping near the event horizon. Hence, the calculations above only consider the silhouette of the invisible black hole shadow (see Ref. \cite{Dokuchaev:2019jqq}), which is prevalent in the literature.

\section{Conclusion} \label{conc}
We derived black hole spacetime metric surrounded with two different dark matter models: (1) apparent dark matter interpreted as a non-local effect of emergent gravity, and (2) an emergent dark matter arising from dark energy viewed as a superconductor. In relation to the dark matter mass distribution, the former is described by the parameter $\xi$ along with a length scale effect from $l$, while the latter is with $\eta$ with the screening length parameter $\lambda_{\rm G}$.

Intrigued by how these two dark matter models affect the black hole geometry at the center of a galaxy, we investigate their effects on the event horizon, null geodesic, and the invisible black hole shadow. We found unique descriptions and differences between the two models, as shown by the plots in Figs. \ref{hor}-\ref{rsh}, which we discussed in Sect. \ref{sec4}. Perhaps one of the intriguing results from this study is that the effect of the parameter $l$ can be considered negligible when applied to SMBHs. Meanwhile, it was shown how this length scale differs from the screening length parameter $\lambda_{\rm G}$ as its effect is more pronounced.

We also found upper bounds to these parameters, which are already depicted in the right panel of Fig. \ref{rsh}. It is a useful plot since it allows us to estimate variables related to $\xi$ and $\eta$. For instance, since the latter is defined as $\xi^2 = (a_0/6)GM$, it has a dimension of $[\text{L}]^2/[\text{t}]^2$. Thus, if we have an upper bound for M87* as $\xi = 0.019$, multiplying by $c^2$ gives $\xi = 1.708 \times 10^{15} \text{ m}^2/\text{s}^2$. The corresponding value for DM mass $M$ is then $\sim 4.91\times 10^{50} \text{ kg} \sim 2.47 \times 10^{20} M_{\odot}$ at $3\sigma$ level. We could also do the same for $\eta$, but $M$ will depend on the specific value for $\alpha$ (or vice-versa). We may have lower estimates for $M$ at lower $\sigma$
levels. In Refs. \cite{Merritt1993, Murphy:2011yz}, the estimated total mass within the halo restricted from the range $40-50$ kpc is around $\sim 6\times 10^{12} M_{\odot}$ at 90\% confidence. While the model considered in this paper does not specify the range of halo concentration, the higher mass $M$ requirement only indicates that a large concentration is needed to affect the shadow. 

The EHT has provided us with confirmation that black holes existed. The cause of measurement uncertainties is yet to be known, perhaps with more sophisticated astronomical instruments in the future (such as the Solar System-based Very Long Baseline Interferometry). It is only then that the imprints coming from certain parameters can be potentially measured, which may confirm which model will be correct. As for now, the study realized that it would take a large mass of dark matter halo to consider the uncertainties from the EHT measurement.

Finally, we will make some remarks on research directions. Eq. \eqref{a(r)} came from the non-approximated dark matter mass, meaning one may also explore dark matter effects on hypothetical micro black holes. That is, to offset the smallness of $l$, one must explore small values of $r$. Another possible avenue of study is exploring or investigating other black hole properties, such as weak and strong deflection angles, quasinormal modes, thermodynamics, and accretion disks. One can also consider the effect of dark matter on wormhole geometries.

\begin{acknowledgements}
R. P. would like to acknowledge networking support of the COST Action CA18108 - Quantum gravity phenomenology in the multi-messenger approach (QG-MM), COST Action CA21106 - COSMIC WISPers in the Dark Universe: Theory, astrophysics and experiments (CosmicWISPers), the COST Action CA22113 - Fundamental challenges in theoretical physics (THEORY-CHALLENGES), and the COST Action CA21136 - Addressing observational tensions in cosmology with systematics and fundamental physics (CosmoVerse).
\end{acknowledgements}

\bibliography{ref}
\end{document}